\documentclass[12pt]{article}
\usepackage{amstext} 
\usepackage{amsmath,amssymb}
\usepackage{theorem} 
\usepackage{enumerate}
\usepackage{tabularx}
\usepackage{graphicx}
\usepackage{sariel,wide}
\usepackage{url,hyperref}

\newcommand{\Wopt}{{\cal W}_{opt}}
\newcommand{\Topt}{{\cal T}_{opt}}

\newcommand{\weight}{{\mathop{\mathrm{weight}}}}
\newcommand{\IDL}{\D_L}
\newcommand{\IDX}[1]{\D_{{#1}}}
\newcommand{\MSTA}{\widehat{\MST}}
\newcommand{\ApproxMST}{{\tt ApproxMST}}
\newcommand{\PropagateWavefront}{{\tt PropagateWavefront}}
\newcommand{\PropagateApproxWavefront}{{\tt PropagateApproxWavefront}}
\newcommand{\cAnother}{c_1}
\newcommand{\cmindist}{c_5}
\newcommand{\cSampleProb}{c_6}
\newcommand{\cFarEnough}{c_7}
\newcommand{\cSample}{c_{samp}}
\newcommand{\Gadj}{G_{adj}} 
\newcommand{\Ot}{\widetilde{O}}
\newcommand{\RS}{{\cal RS}}
\newcommand{\polylog}{\mathop{\mathrm{polylog}}}
\newcommand{\vL}{\vec{v}_L}
\newcommand{\plow}{z}
\newcommand{\phigh}{Z}
\newcommand{\HX}{{\cal H}}
\newcommand{\R}{{\cal R}}
\newcommand{\lshort}{l_{short}}
\newcommand{\out}{\mathrm{out}}

\newcommand{\EmbedDim}{7}

\title{When Crossings Count --- Approximating the Minimum
   Spanning Tree\thanks{A preliminary version of the paper
      appeared in the {\em 16th ACM Symposium of
         Computational Geometry}, 166--175, 2000.}}
\author{Sariel Har-Peled\sarielthanks{}
   \and
   Piotr Indyk\thanks{MIT Laboratory for Computer Science;
         545 Technology Square, NE43-373;
         Cambridge, Massachusetts 02139-3594;
         {{\tt indyk\atgen{}theory.lcs.mit.edu}}}}
\date{\today}

\begin{document}
%\let\ps@plain=\ps@empty
%\nopagenumber{}
\maketitle
%\renewcommand{\thefootnote}{}
%\copyrightspace{}
%\renewcommand\thefootnote{\arabic{footnote}}%
\begin{abstract}
    We present an $(1+\eps)$-approximation algorithm for
    computing the minimum-spanning tree of points in a
    planar arrangement of lines, where the metric is the
    number of crossings between the spanning tree and the
    lines. The expected running time of the algorithm is
    near linear. We also show how to embed such a crossing
    metric of hyperplanes in $d$-dimensions, in subquadratic
    time, into high-dimensions so that the distances are
    preserved.  As a result, we can deploy a large
    collection of subquadratic approximations algorithms
    \cite{im-anntr-98,giv-rahdp-01} for problems involving
    points with the crossing metric as a distance function.
    Applications include MST, matching, clustering,
    nearest-neighbor, and furthest-neighbor.
\end{abstract}

\begin{figure}
    \centerline{\includegraphics{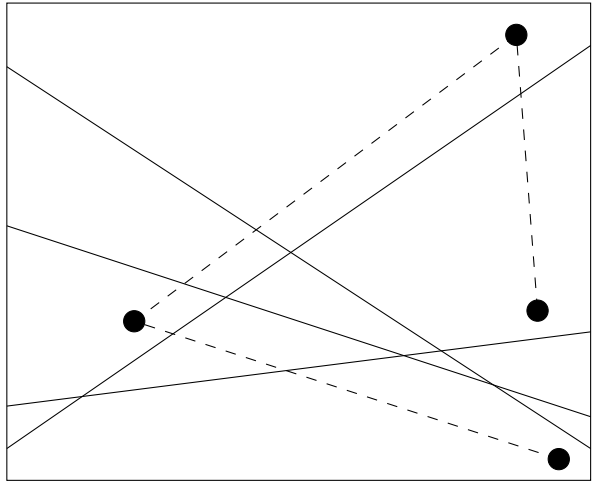}}

    \caption{A set of lines and points, and the resulting
       crossing MST. Note that in this case the crossing MST
       is different from the Euclidean MST.}
    
    \figlab{mst}
\end{figure}

%%%%%%%%%%%%%%%%%%%%%%%%%%%%%%%%%%%%%%%%%%%%%%%%%%%%%%%%%%%%%%%%%%
%%%%%%%%%%%%%%%%%%%%%%%%%%%%%%%%%%%%%%%%%%%%%%%%%%%%%%%%%%%%%%%%%%

\section{Introduction}

Given a set of lines in the plane a natural measure of
distances between any two points is the number of lines one
has to cross to reach from one point to the other.  This is
a discrete distance measure that can be used to approximate
the Euclidean distance and other distance measures.
However, since this measure is defined by an arrangement of
lines it is not locally defined and is thus computationally
cumbersome. Finding the minimum spanning tree (MST) of a set
of points, so that the number of intersections between the
tree and the given set of lines is minimized, quantify how
the set of points interact with the set of lines; see Figure
\ref{fig:mst}. In fact, when the set of lines is the set of
all possible lines, then this MST is the standard Euclidean
MST \cite{af-acrsm-97} (here one minimizes the average
number of edges of the MST crossed when picking a random
line).  Such an MST is related to a spanning tree of low
stabbing number (STLSN) \cite{w-stlcn-92,a-idapa-91}. While
the spanning tree of low stabbing number guarantee that any
line intersects at most $O(\sqrt{n})$ edges of the spanning
tree, the MST guarantees that the overall number of
intersections between the tree and a given set of lines is
minimized.  Thus, if we have the set of lines in advance
then the MST will have overall less intersections than the
STLSN.  The spanning tree of low-stabbing number was used in
several applications, see for example
\cite{a-idapa-91,mww-deabv-91}.  In particular, having such
an MST enables one: (i) to answer half-plane range queries
in an efficient manner using a near linear space
\cite{ghs-citds-91}, (ii) bound the complexity of the faces
of the arrangement of lines that contain the points
\cite{hs-oplpa-01-dcg}, and (iii) traverse between the points in
an efficient way, so that the number of updates needed is
minimized. (Imagine traversing among the points and
maintaining the set of half-planes that contain the current
point. Each time one crosses a line an update operation is
performed.)

Computing the MST for the general case of arcs can be done
in $O(n^2\log{n})$ time by performing wavefront propagation
from each of the points (see Section
\ref{sec:cont:dijkstra}). As for approximation algorithms,
Har-Peled and Sharir \cite{hs-oplpa-01-dcg} gave recently an
approximation algorithm for the case of arcs, computing a
Steiner tree in expected running time $w=O(
\lambda_{t+2}(n+\Wopt) \log(n))$, where $t$ is the maximum
number of intersections between a pair of arcs,
$\lambda_{t+2}(\cdot)$ is the maximum length of a
Davenport-Schinzel sequence of order $t$, and $\Wopt$ is the
weight of the optimal Steiner tree.\footnote{It is easy to
   verify that if we have triangle inequality then the
   Steiner tree weight is at least half the weight of the
   MST.} The algorithm outputs a tree of weight $w$ (and
thus gives roughly $O(\log n)$-approximation).

In this paper, we present two results:
\begin{itemize}
    \item A near linear time $(1+\eps)$-approximation algorithm to the
    minimum-spanning tree under the crossing metric in the
    planar case. 
    
    \item We show to embed the crossing metric among
    hyperplanes into a Hamming distance in high dimensions.
    As a result, we show how one can apply known
    subquadratic approximation algorithms for problems
    involving point-sets and hyperplanes in high dimensions
    (MST, clustering, matching, etc).  
    
    \remove{ Intuitively, all those applications rely on a
       black-box (called $\eps$-PLEB in \cite{im-anntr-98})
       that decides whether, for a query point $q$, there is
       no point close to $q$ in our point-set (i.e., $\geq
       r$ for all points), or alternatively finds a nearby
       point (i.e., $\leq (1+\eps)r$) under the crossing
       metric, where $r$ is a prespecified threshold.
       
       Using $\eps$-PLEB for the embedded points
       \cite{im-anntr-98} we construct the required
       $\eps$-PLEB for the points we started with.  For
       $d>2$ dimensions, we show how to embed the crossing
       metric induced by $n$ hyperplanes over a set of $n$
       points in $\Re^d$, in $O(n^{2d/(d+1) + \delta})$
       time, where $\delta>0$ is arbitrary.
    
       we can maintain the dynamic $c$-approximate nearest
       neighbor problem over this $n$-point metric in
       $\Ot( n^{1/c})$ time per
       operation~\cite{im-anntr-98}.  This in turn implies
       dynamic amortized $\Ot(n^{4/3} +
       n^{1+1/c})$-time $c$-approximation algorithms for
       bichromatic closest pair~\cite{e-fhcoa-98} and
       $\Ot(n^{4/3} + n^{1+1/c})$-time algorithms for:
       $c$-approximate diameter and discrete minimum
       enclosing ball \cite{giv-rahdp-01},
       $O(c)$-approximate facility location and bottleneck
       matching (all for $d=2$) \cite{giv-rahdp-01}.  }
 
    The connection between the crossing metric, and points
    in high dimension follows by interpreting the input
    points as points in abstract VC-space \cite{pa-cg-95}
    induced by the lines. Namely, we associate with each
    point in the plane, an $n$-dimensional binary vector,
    where $i$-th coordinate indicate on which side of the
    $i$-th line the point lies.  In this way, we mapped our
    input points into points lying on the $n$-dimensional
    hypercube. The crossing metric is no more than the
    Hamming distance between the mapped points.  We can now
    deploy the techniques of \cite{im-anntr-98} to those
    mapped points, yielding an approximation algorithm for
    the MST problem.  Bringing down the running time to be
    subquadratic requires some additional work.
    
    Specifically, we show how to compute a mapping of the
    points into space of dimension $O(\log^\EmbedDim n)$; this
    embedding can be computed in $\Ot(n^{4/3})$
    time\footnote{Here and in the rest of this paper
       $f(n)=\Ot(g(n))$ iff $f(n)=g(n)
       (1/\eps)^{O(1)}\log^{O(1)}n$, and
       $f(n)={O_\eps}(g(n))$ iff $f(n)=O( g(n)
       /\eps^{O(1)})$}, for $n$ points, so that we get a
    $(1+\eps)$ gap property for a specified range of
    distances is preserved.
    
    As a result, we can solve several approximation problems
    for this metric, among them is the MST problem.  In
    fact, our near-linear approximate MST algorithm in the
    plane can be roughly viewed as an unraveling of the
    corresponding MST approximation algorithm in high
    dimensions. Similar bounds can be derived for $d > 2$
    dimensions. See \secref{embed} for details.
\end{itemize}

The paper is organized as follows: In Section
\ref{sec:cont:dijkstra}, we describe how one can compute the
exact MST using wavefront propagation.  In \secref{speedup},
we present the planar $(1+\eps)$-approximation algorithm for
the MST.  Next, in Section \ref{sec:embed}, we
describe the embedding into points in high dimension and
demonstrate its usage for computing an approximate MST.
Concluding remarks are given in Section \ref{sec:conc}.

\begin{figure}
    \begin{center}
        \begin{tabular}{cc}
            {\includegraphics{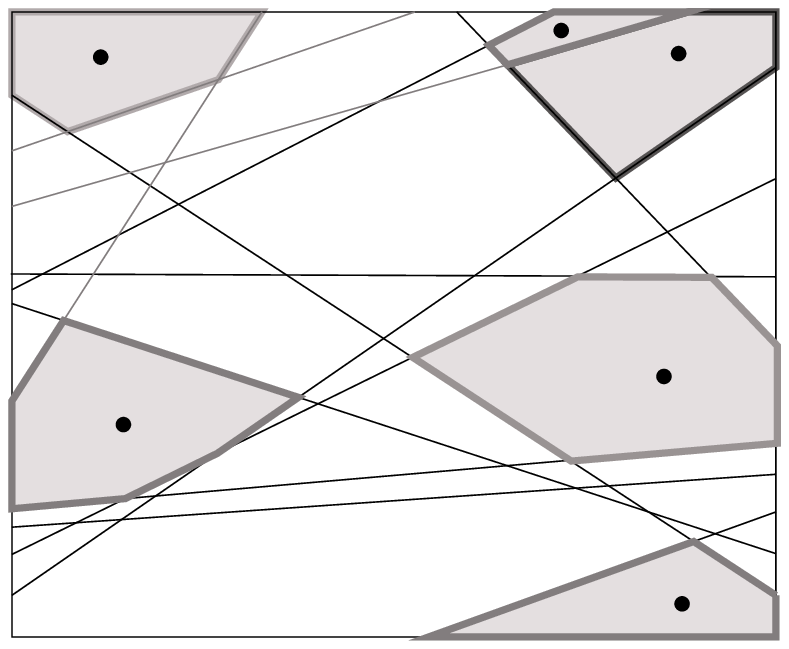}}
            &
            {\includegraphics{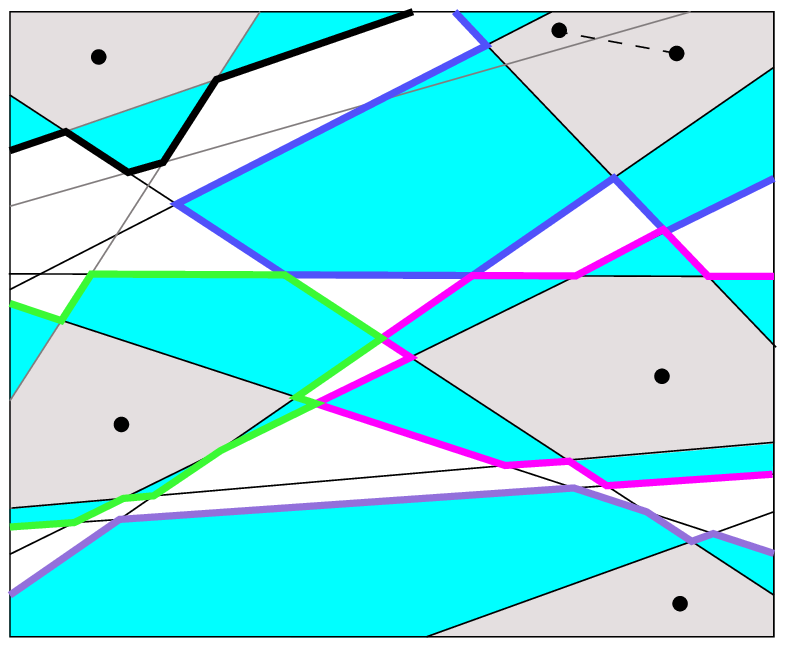}}\\
            (i) & (ii)\\
            \\
            {\includegraphics{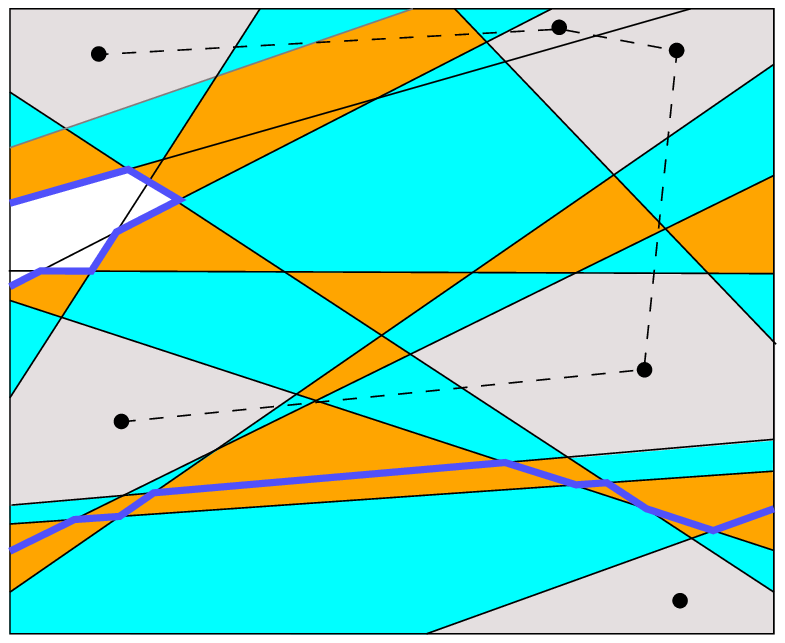}}
            &
            {\includegraphics{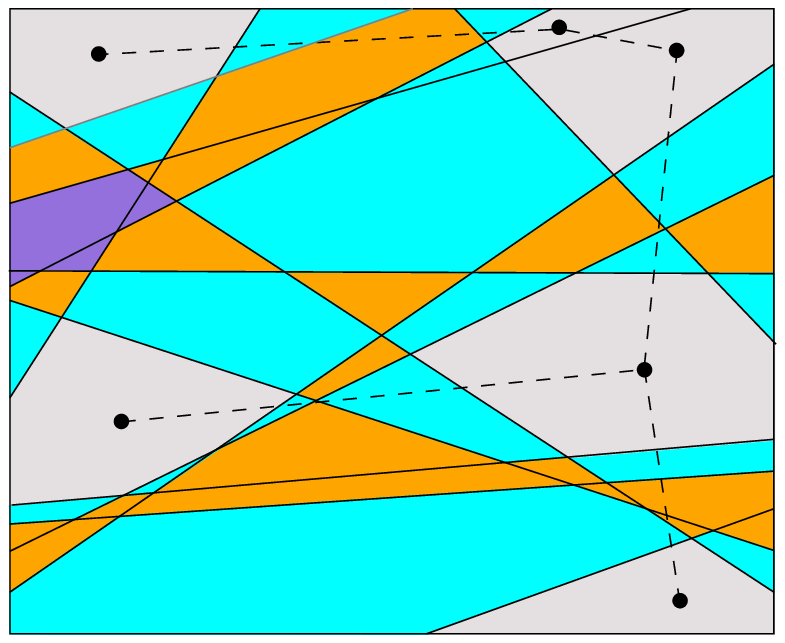}}\\
            (iii) & (iv)\\
        \end{tabular}
    \end{center}
    
    \caption{Computing the crossing MST by doing wavefront
            propagation. The thick lines denote the boundary
            of the current connected components of the
            spanning forest.}
  
    \figlab{mst:wavefront}
\end{figure}

%-------------------------------------------------------------
%-------------------------------------------------------------

\section{Minimum Spanning Tree by Continuous Dijkstra}
\seclab{cont:dijkstra}

In this section, we present a simple algorithm for computing
the crossing MST. It relies on a simple direct solution
interpreted as a geometric algorithm. We also present a
``weight sensitive'' algorithm (\lemref{propagate}) that
computes portions of the MST in time proportional to its
overall weight.

In the following, we assume that we are given a set $L$ of
lines and a set $P$ of points in the plane. For simplicity,
we assume $|P| = |L| = n$.

\begin{defn}
    For a set $L$ of lines, the {\em crossing metric} is
    defined to be the minimum number of lines of $L$ that
    one has to cross as one moves between two prespecified
    points. Thus, for a pair of points $p,q \in \Re^2$ the
    crossing distance between $p$ and $q$, denoted by
    $\IDL(p,q)$, is the number of lines of $L$ that
    intersects the segment $p q$.  If $L$ is a set of arcs, a
    similar crossing metric is defined, although the
    ``shortest path'' in this case is no longer necessarily
    a straight segment.
\end{defn}

\begin{defn}
    For a set $L$ of lines, and a set $P$ of points in the
    plane, let $\Topt(P,L)$ denote a minimum spanning tree
    of $P$ under the crossing metric induced by $L$, and let
    $\Wopt(P,L)$ denote the weight of $\Topt(P,L)$.
\end{defn}

Let $\Arr = \Arr(L)$ denote the planar arrangement induced
by the lines of $L$.  Let $\Gadj = \Gadj(\A)$ be the
adjacency graph of $\Arr$; namely, each face of $\Arr$ is a
vertex, and two vertices are connected if the two
corresponding faces share an edge.  Let $V$ be the set of
vertices of $\Gadj$ that corresponds to the faces of $\Arr$
that contains points of $P$. Clearly, the crossing MST of
$P$ in $\Arr$, corresponds to the MST of $V$ in the graph
$\Gadj$ (here, each edge has associated weight $1$).

Computing the MST of $V$ in $\Gadj$ can be done by performing
a simultaneous flooding of $\Gadj$ from the vertices of $V$.
Indeed, we compute in the $i$-th iteration all the vertices
of $\Gadj$ that are in distance $\leq i$ from any vertex of
$V$.  This can be easily done using a modified BFS. In the
beginning, the flood front is made out of $n$ connected
components.  Every time two connected components of the flood
front collide, we discovered a new edge of the MST. This edge
connects the two vertices that induced the two parts of the
wavefront that collided. This is a somewhat non-standard
algorithm for computing the MST, but one can easily verify
that it indeed computes the MST of $V$ in $\Gadj$.

This flooding algorithm has a natural geometric
interpretation: Let $\F_{2i}$ denote the set of all faces of
$\Arr$ that are in (crossing) distance at most $i$ from any
point of $P$.  Clearly, $\F_0$ is the set of faces of $\Arr$
that contain points of $P$.  The algorithm works in $n/2$
phases. We do a wavefront propagation in $\Gadj$, starting
from all the vertices that correspond to the marked faces
(i.e., faces of $\Arr$ that contain points of $P$). In each
iteration, we propagate the wavefront from the faces of
$\F_{2i-2}$ into the faces of $\F_{2i}$.  It is easy to
verify that a connected component of the flood corresponds
to a connected component of the wavefront of $\F_{2i}$.
(Note, that two faces of $\F_{2i}$ might be adjacent but
belong to different wavefronts as the wavefronts did not
cross the separating edge yet and thus were not merged into
a single wavefront.)  The connected components are
maintained implicitly by a union-find data-structure. In
particular, during the $i$-th iteration of the wavefront
propagation in $\Gadj$, when two different connected
components of the wavefront collide, it corresponds to two
points of $P$ with crossing distance equal to $2i-1$ or $2i$
from each other.

In particular, if there is an edge of the MST of weight
$2i-1$ or $2i$ it would be discovered when the corresponding
wavefronts collide.  The $i$-th iteration of the wavefront
propagation, corresponds to the detection of edges of weight
$2i-1$ and $2i$ in the MST. For the MST applications, we
first handle all relevant edges of weight $2i-1$, and later
all such edges of weight $2i$.  This requires a somewhat
careful implementation, and we omit the the technical but
straightforward details. See Figure \ref{fig:mst:wavefront}.
 
Note, that the wavefront propagation can be done without
constructing $\Gadj$ in advance, and one can compute parts
of $\Gadj$ on the fly as needed (i.e., we need to compute
only the parts of $\Gadj$ that are covered by the wavefront,
or are about to be covered). Of course, in the worst case,
the whole graph $\Gadj$ would be computed, which takes
$O(n^2 \log{n})$ time (this corresponds to computing the
whole arrangement $\Arr(L)$).

\begin{lemma}
    Given a set $L$ of $n$ lines, and a set $P$ of $n$ points, a
    minimum spanning tree $\Topt(P,L)$ of $P$ under the crossing
    metric $\IDL$ can be computed in $O(n^2\log{n})$ time.

    \lemlab{bf:prop}
\end{lemma}

\begin{remark}
    In the algorithm of \lemref{bf:prop} we did not use the fact
    that $L$ is a set of lines. The same algorithm will work for the
    case where $L$ is a set of arcs. Since we do not have the triangle
    inequality in this case, the edges of the MST are no longer line
    segments, but rather a Jordan arcs. (For example, imagine that the
    set $L$ is a single segment and we would like to connect two
    points that are separated by this segment. This can be done with
    no crossing by going ``around'' this segment.)
\end{remark}

To be able to generate parts of $\Gadj$ incrementally, as we
perform the wavefront propagation, we need a way to compute
the relevant portions of $\Arr(L)$ on the fly.

\begin{theorem}[\cite{hs-oplpa-01-dcg}]
    Let $L$ be a set of $n$ lines, as above, and $P$ a set
    of $m$ points in the plane. Then one can compute, in
    expected $O(\pth{ n+w +m}\alpha(n)\log{n})$ time,
    a Steiner tree $\MSTA$ of $P$, so that the expected
    weight of $\MSTA$ is $O((n+w)\alpha(n)\log{n})$,
    where $w= \Wopt(P,L)$ and $\alpha(n)$ is the inverse of
    the Ackermann function. Alternatively, one can compute
    the $m$ faces that contain the points of $P$ in the same
    time bound.

    \theolab{hs}
\end{theorem}

\begin{lemma}[\cite{a-idapa-91}]
    There exists a Steiner tree $\MST'$ of $P$, so that
    $\Wopt(P,L) = O(n\sqrt{n})$, and this is tight in the
    worst case (even for the case the arcs are lines).
    \lemlab{span:tree}
\end{lemma}

In the worst case, Theorem \ref{theo:hs} is inferior to
implicit point-location data-structures \cite{ams-cmfal-98}
(which can perform the implicit point-location needed in
roughly $O(n^{4/3})$ time for $m=n$), as implied by
\lemref{span:tree} (as the weight of the MST is
$\Omega(n^{3/2})$ in the worst case, and this is the time to
compute the relevant portions of the arrangement using the
algorithm of \theoref{hs}).  However, the running time of
the algorithm of \theoref{hs} is sensitive to the overall
weight of the MST. This would be crucial for our algorithm.

\begin{lemma}
    Given a set $L$ of $n$ lines, a set $P$ of $n$ points,
    and a parameter $i$, one can compute, in expected
    $O(i(n+\Wopt)\alpha^2(n)\log{n})$ time, a minimum
    spanning forest of $P$ under the crossing metric $\IDL$,
    that connects all the points of $P$ in distance at most
    $\leq 2i$ from each other, where $\Wopt = \Wopt(P,L)$.

    \lemlab{propagate}
\end{lemma}

\begin{proof}
    The wavefront propagation on $\Gadj$ can be done using
    an implicit representation of the arrangement of
    $\Arr(L)$.  Namely, we compute the set $\F_i$ of faces
    of $\Arr(L)$ in distance $i$ from the points of $P$.
    Observe that the complexity of $\F_i$ is $O((n+\Wopt)
    i\alpha( n/i))$. Indeed, the points of $P$ can be
    connected by an arc $\gamma = \Topt(P,L)$ having
    $O(\Wopt)$ intersections with the lines of $L$, and let
    $\Arr'$ be the arrangement resulting from $\Arr$ by
    creating a tiny gate for each intersection of $\gamma$
    with the lines of $L$. The zone of $\gamma$ in $\Arr(L)$
    corresponds to a single face $F$ of $\Arr'$, and the
    faces of $\F_i$ are contained in the set of faces in
    distance $\leq i$ from $F$. By \cite{bds-lric-95}, the
    complexity of this region is $O((n+\Wopt) i\alpha(
    n/i))$ (this is a bound on the complexity of all the
    vertices in distance $\leq i$ from the face $F$.).
    
    Clearly, the faces of $\F_i$ have a spanning tree of
    weight $O((n+\Wopt) i\alpha( n/i))$, and so it can be
    computed in an online fashion in $O((n+\Wopt) i\alpha^2
    ( n) \log{n})$ expected time, by Theorem \ref{theo:hs}.
\end{proof}

\begin{figure*}[tb]
    \vspace{-0.5cm}
    \begin{center}
        \fbox{
           \begin{program}
               \> \>{\large{\sc{Algorithm}}}\ \ \ 
               \Proc{\ApproxMST{}($P, L, \eps$)} \\
               \> \>{\tt Input:} {\rm{A set of points $P$, 
                     a set of lines $L$, and an approximation
                     parameter $\eps$}}\\
               \> \>{\tt Output:} A spanning tree of $P$ of
               weight $\leq (1+\eps)\Wopt(P,L)$\\
               \> \Procbegin \\
               \>\> $M \leftarrow $ Approximate the weight
               of MST
               using the algorithm of Lemma \lemref{rough}.\\
               \> \> $l_0 \leftarrow \max \pth{ \frac{\eps
                     M}{\cmindist n \alpha(n)\log^2{n}}, 1}
               $\\
               \> \> Set $F = (P, \emptyset)$ to be the
               an empty spanning forest of $P$.\\

               \>\>\PropagateApproxWavefront{}( $P$, $L$,
               $l$, $F$ )\\

               \>\> $i \leftarrow 1$\\
               \> \> \While $F$ is not a single connected
               component \Do\\ 
               \>\>\> $l_i \leftarrow l_{i-1} \cdot 2$\\               
               \>\>\>\PropagateApproxWavefront{}( $P$, $L$,
               $l$, $F$ )\\
               \>\>\> $i \leftarrow i + 1$\\
               \> \> \End \While\\
               \>\>\\ 
               \>\>\Return $F$ \\
               \>\Endproc{\ApproxMST{}}
           \end{program}
        }
    \end{center}
    \vspace{-0.5cm}
    \caption{Approximating the MST in the Plane}
    \figlab{alg:mst2}
%    \vspace{-0.5cm}
\end{figure*}

\begin{figure*}[tb]
    \vspace{-0.5cm}
    \begin{center}
        \fbox{
        \begin{program}
            \> \>{\large{\sc{Algorithm}}}\ \ \ 
            \Proc{\PropagateWavefront{}( $P$, $R$, $l$, $F$ )}\\
            \> \>{\tt Input:} ~
            {\rm{$P$ -  set of points}}\\
            \>\>\>\>\>{\rm{$R$ - set of lines}}\\
            \>\>\>\>\>{\rm{$l$ - propagation distance}}\\
            \>\>\>\>\>{\rm{$F$ - current spanning forest}}\\
            \> \>{\tt Output:} An updated forest $F$ with
            any pair of points of distance $\leq 2l$ in a\\
            \>\>\>\>\> single
            connected component\\            
            \> \Procbegin \\
            \>\> Initialize the data-structure $D(R)$ of
            \cite{hs-oplpa-01-dcg} for online point-location.\\ 
            \> \> Set $W_0$ to be the set of faces of
            $\Arr(R)$ that contains points of $P$.\\
            \>\>\>\>Use
            $D(R)$ to compute those faces.\\
            \> \> \For $i=1, \ldots, l$ \Do\\
            \>\>\> $W_i \leftarrow$ 
            Set of faces of
            $\Arr(R)$ of distance $=i$
            from points of $P$.\\
            \>\>\>\>\>Do wavefront propagation from
            $W_{i-1}$, and use $D(R)$ to retrieve
            \\
            \>\>\>\>\>the faces of interest in $\Arr(R)$. \\ 
            \>\>\> \If two different wavefronts collide
            \Then\\
            \>\>\>\> Add
            an edge connecting the two corresponding points
            to $F$\\ 
            \>\>\>\>Merge the corresponding connected
            components.\\
            \>\> \Endfor\\
            
            \>\Endproc{\PropagateWavefront{}}
        \end{program}
        }
    \end{center}
    \vspace{-0.5cm}
    \caption{Doing the wavefront propagation}
    \figlab{alg:propagate}
    \vspace{0.5cm}
\end{figure*}

\begin{figure*}[tb]
    \vspace{-0.5cm}
    \begin{center}
        \fbox{
        \begin{program}
            \> \>{\large{\sc{Algorithm}}}\ \ \ 
            \Proc{\PropagateApproxWavefront{}( $P$, $L$, $l$, $F$ )}\\
            \> \>{\tt Input:} ~
            {\rm{$P$ -  set of points}}\\
            \>\>\>\>\>{\rm{$L$ - set of lines}}\\
            \>\>\>\>\>{\rm{$l$ - starting propagation distance}}\\
            \>\>\>\>\>{\rm{$F$ - current spanning forest}}\\
            \> \>{\tt Output:} An updated forest $F$ with
            any pair of points of distance $\leq 2l$ in a\\
            \>\>\>\>\> single
            connected component\\            
            \> \Procbegin \\
            \>\>\> Compute a random sample $R$ by choosing
            each line of $L$ into the sample with
            \\
            \>\>\> \> \> probability $f(l) = 128
            \cSampleProb
            \frac{\log{n}}{l\eps^2}$\\
            \> \> \> {\tt /* Approximate the wavefront propagation
               in $A(L)$ by doing}\\
            \>\>\>\> {\tt it (exactly) in $\Arr(R)$ */}\\
            \> \> \> \PropagateWavefront{}( $P$, $R$, $\cFarEnough
            \log{n}/\eps^2$, $F$ )\\
            \>\>\>\>\>\>{\tt /* 
            $\cFarEnough$  is an appropriate constant */}\\
            \>\Endproc{\PropagateApproxWavefront{}}
        \end{program}
        }
    \end{center}
    \vspace{-0.5cm}
    \caption{Doing the approximate wavefront propagation}
    \figlab{alg:propagate:x}
    \vspace{0.5cm}
\end{figure*}

%-------------------------------------------------------
%-------------------------------------------------------
\section{Approximation Algorithm for the Planar Case}
\seclab{speedup}

The algorithm is depicted in \figref{alg:mst2},
\figref{alg:propagate} and \figref{alg:propagate:x}. We next
describe the algorithm and its analysis in more detail.

\lemref{propagate} provides us with an algorithm for
approximating the MST in roughly quadratic time in the worst
case.  To get a near linear running time, we simulate the
Dijkstra algorithm by performing the wavefront propagation
in an approximate fashion.

\begin{defn}
    A metric $\D'$ {\em $\eps$-approximates} a metric $\D$,
    if for any $p,q,r,s \in P$ such that $\D'(p,q) \leq
    \D'(r,s)$ then $\D(p,q) \leq (1+\eps)\D(r,s)$.
\end{defn}

\begin{defn}
    For a set $F$ of segments in the plane, and a metric $\D$,
    let $\weight_\D(F) = \sum_{e \in F} \D(e)$ denote the
    total weight of $F$ under the metric $D$.
\end{defn}

The proof of the following lemma is straightforward, and is
included only for the sake of completeness.
\begin{lemma}
    Let the metric $\D'$ be an $\eps$-approximation to the
    metric $\D$ over a point-set $P$. Let $T'$ be an MST of
    $P$ under $\D'$. Then, $\weight_\D(T') \leq
    (1+\eps)\weight_\D(T)$, where $T$ is the MST of $P$
    under $\D$, and $\weight(T)$ is the total weight of the
    edges of $T$.

    \lemlab{approx:mst}
\end{lemma}

\begin{proof}
    Let $e_1', \ldots, e_{n-1}'$ be the the edges of $T'$
    sorted by their weight $\D'(e_1') \leq \ldots \leq
    \D'(e_{n-1}')$. Let $T_0 = T$, and let $T_i$ be the tree
    resulting from removing the heaviest edge (according to
    $\D'$) from the cycle present in $T_{i-1} \cup
    \brc{e_i'}$ (if $e_i'$ is already in $T_{i-1}$ we do
    nothing). Let $e_i$ denote this removed edge. Clearly,
    $\D'(e_i') \leq \D'(e_i)$ and, by definition, $\D(e_i')
    \leq (1+\eps) \D(e_i)$. Namely, we replaced an edge
    $e_i$ by an edge $e_i'$ which is heavier by a factor of
    $(1+\eps)$. In the end of the process $T_{n-1}$ is just
    $T'$, and $\weight_\D(T') \leq
    \sum_{i=1}^{n-1}(1+\eps)\weight_\D(e_i) \leq
    (1+\eps)\weight_\D(T)$.
\end{proof}

\lemref{approx:mst} suggest that if we can find a
computationally cheaper approximate metric than
$\IDL(\cdot,\cdot)$, then we can use it to compute the MST.
A natural way to do that, is to randomly sample a subset $R
\subseteq L$, and use $\IDX{R}( \cdot, \cdot )$ as the
approximate metric. However, it is easy to verify that
$\IDX{R}$ is an $\eps$-approximate metric to $\IDL$ only if
$L = R$.

\begin{defn}
    Let $\D',\D$ be two metrics, $\eps > 0$, and $l$ be
    prescribed parameters.  The metric $\D'$ is an {\em
       $(\eps,l)$-approximation} to $\D$, if for any
    $p,q,r,s \in P$, such that (i) $\D( p,q), \D(r,s) \geq
    l$, and (ii) $\D'(p,q) \leq \D'(r,s)$, we have $\D(p,q)
    \leq (1+\eps)\D(r,s)$.
    
    Namely, $\D'$ $\eps$-approximates $\D$ for distances not
    smaller than $l$.
\end{defn}

\begin{defn}
    For $l, \eps$, let $\nu(l, \eps ) = \max \pth{ 128
       \cSample \frac{\log{n}}{l\eps^2}, 1 }$, where
    $\cSample$ is an appropriate constant. Let $\RS( L, l,
    \eps)$ be a random subset of $L$ generated by picking
    independently each line of $L$ with probability
    $\nu(l,\eps)$.
    
    Let $\rho(l, \eps) = \nu(l, \eps) l = 128 \cSample
    \frac{\log{n}}{\eps^2}$. The value $\rho(l,\eps)$ is the
    expected crossing distance in $\Arr(\RS(L, l, \eps))$
    between two points $p, q \in P$ such that $\IDL(p,q) =
    l$.
    \deflab{def:sample}
\end{defn}

\begin{lemma}
    Let $L$ be a set of $n$ lines in the plane, $l$ a
    positive integer number, $\eps >0$, and let $R = \RS(L,
    l, \eps)$ be a random subset of $L$.
    
    For any two points $p,q$ of distance $\IDL(p,q) \geq l$
    from each other we have
    \[
    \IDL(p,q) \leq \frac{n}{r(1-\eps/4)}\cdot \IDX{R}(p,q)
    \leq (1+\eps)\IDL(p,q),
    \]
    with probability $\geq 1-n^{-c_0}$.

    Furthermore, $\IDX{R}( \cdot, \cdot)$ is an
    $(\eps,l)$-approximation to $\IDL(\cdot, \cdot)$ with
    high probability.

    \lemlab{good:estimate}
\end{lemma}

\begin{proof}
    Indeed, let $X_{p q} = D_R(p,q)$. We have,
    \begin{eqnarray*}
        \mu = E[ X_{p q} ] = \IDL(p,q)\cdot \nu( l, \eps ) 
        \leq 128 \IDL(p,q) \cSample \frac{\log{n}}{l \eps^2}
        \geq \frac{128 \cSample
           \log{n}}{\eps^2}.
    \end{eqnarray*}

    By Chernoff inequality \cite{mr-ra-95,mps-lpvaa-98}, we have that
    \begin{eqnarray*}
        P \pbrc{ \cardin{X_{p q} - \mu} > \frac{\eps}{4}\mu} &\leq& 2
        \pth{ \frac{e^{\eps/4}}{{\pth{1 + \frac{\eps}{4}}^{1+
                    \eps/4}}}}^\mu
        = 2 \exp \pth{\mu \pth{ \frac{\eps}{4} -
              \pth{1+\frac{\eps}{4}}\log \pth{ 1 + \frac{\eps}{4}}}} \\
        &\leq& 2 \exp \pth{\mu \pth{
              \frac{\eps}{4} - \pth{1+\frac{\eps}{4}}
              \pth{ \frac{\eps}{4} - \frac{\eps^2}{32}}}}\\
        &\leq& 2
        \exp \pth{-\mu \frac{\eps^2}{64}}
        \leq
        \exp \pth{- \frac{128 \cSample
              \log{n}}{\eps^2} \cdot \frac{\eps^2}{64}}
        \leq n^{-\cSample},
    \end{eqnarray*}
    since $\log(1+x) \geq x - x^2/2$, for $0 \leq x \leq 1$. In
    particular, this implies that with high probability
    $\mu(1-\eps/4) \leq X_{p q} \leq \mu
    (1+\eps/4)$. Namely, with high probability we have
    \begin{eqnarray*}
        \IDL(p,q) &\leq& \frac{X_{p q}}{\nu(l, \eps )(1-\eps/4)} 
        \leq
        \frac{\nu(l, \eps )(1+\eps/4)}{\nu(l, \eps )(1-\eps/4)} \IDL(p,q)
        =
        \frac{1+\eps/4}{1-\eps/4} \IDL(p,q) \\
        &\leq &
        (1+\eps)\IDL(p,q).
    \end{eqnarray*}

    Consider now four points $p,q,r,s$, such that
    $\IDL(p,q), \IDL(s,t) \geq l$ and $\IDX{R}(p,q) \leq
    \IDX{R}(r,s)$. By the above discussion, we have with
    high probability
    \[
    \IDL(p,q) \cdot \nu(l,\eps) (1-\eps/4) \leq \IDX{R}(p,q)
    \leq \IDX{R}(r,s) \leq (1+\eps) \IDL(r,s) \cdot
    \nu(l,\eps)(1-\eps/4).
    \]
    Namely, $\IDL(p,q) \leq (1+\eps)\IDL(r,s)$. Namely,
    $\IDX{R}(\cdot, \cdot)$ is an $(\eps,l)$-approximation
    to $\IDL(\cdot, \cdot)$ with probability $\geq 1 - {n
       \choose 2} n^{-\cSample}$.
\end{proof}

\lemref{good:estimate} and \lemref{approx:mst} suggest that
we compute the MST by computing an appropriate random sample
$R$ (by using a threshold $l$), and deploy the algorithms of
\secref{cont:dijkstra} to compute the MST of $P$ in
$\Arr(R)$. Such an MST would be an approximate MST. There
are two main problems with this approach: (i) For short
distances (i.e., $l=1$), just starting the wavefront
propagation (i.e., \lemref{propagate}) is prohibitively
expensive (it roughly takes $O(\Wopt(P,L))$ time which might
be $\Omega(n^{3/2})$), (ii) For long distances (i.e., $\geq
i \cdot l$), the wavefront propagation becomes, again,
prohibitly expensive (i.e. $\Ot(ni)$) by
\lemref{propagate}.

\begin{corollary}
    Let $U$ be the total weight of all the edges of $\T$
    having weigh less than $\eps\Wopt(P,L)/(10n)$. Then $U
    \leq \eps\Wopt(P,L)/10$.
    \corlab{idiotic}
\end{corollary}

\lemref{rough} describes how we can approximate $\Wopt(P,L)$
to within a polylogarithmic factor using random sampling in
near linear time.  Since the algorithm of this lemma is very
similar to the techniques used below, we defer its
description to the appendix.  Equipped with such
approximation $M$, we know by \corref{idiotic} that we do
not ``care'' about edges of the MST of length smaller than
$l_0 = O(\eps M/(n \polylog(n)))$.  In particular, we can
generate a random sample $R_0$ which provides an
$(\eps,l_0)$-approximation to $\IDL(\cdot, \cdot)$. Thus, we
can approximate the MST by computing the MST of
$\Topt(P,R_0)$.

This, however, does not address the second problem. Indeed,
computing the MST of $\Topt(P,R_0)$ might still be too
expensive, as the following lemma testifies.

\begin{lemma}
    Given a set $L$ of $n$ lines, a set $P$ of $n$ points,
    and parameters $l, i, \eps, U$, such that $l
    =\Omega\pth{ \Wopt(P,L)/(n U) }$ and let $R =\RS(L, l,
    \eps)$ be a random sample of $L$.  Then, one can
    compute, in expected $\Ot (i U n)$ time, a minimum
    spanning forest of $P$ under the crossing metric
    $\IDX{R}$, that connects all the points of $P$ in
    distance at most $\leq 2i$ from each other.
    
    \lemlab{propagate:ext}
\end{lemma}

\begin{proof}
    Let $X$ denote the size of $R$. Clearly, The expected
    value of $X$ is
    \[
    E[X] = n \nu(l, \eps) = 128 n \cSample
    \frac{\log{n}}{l\eps^2} = O \pth{ \frac{U n^2 \log
          n}{\eps^2\Wopt(P,L)}},
    \]
    by \defref{def:sample}.  Let $\gamma = \Topt(\gamma,L)$.
    Let $Y = \weight(\gamma, R)$. Clearly,
    \[
    E[Y] = \weight(P,R) = \Wopt(P,L) \nu(l, \eps) = O \pth{
       \frac{U n \log n}{\eps^2}}.
    \]
    Namely, $E[ \Wopt(P,R) ] \leq E[ Y] = O \pth{ \frac{ U n
          \log n}{\eps^2}}$.  The running time bound now
    follows immediately by applying the algorithm of
    \lemref{propagate} to $P$ and $R$.
\end{proof}

The algorithm of \lemref{propagate:ext} first performs
wavefront propagation for distances in $\Arr(R)$ which are
smaller than $\rho( l, \eps)$.  For such distances $\Arr(R)$
{\em does not provide} reliable estimate (i.e., ordering) of
the crossing distances between points.  However, once the
distances propagated exceed $\rho(l,\eps)$, we know by
\lemref{good:estimate} that the distances are now
$(\eps,l)$-approximated correctly. The main importance of
the algorithm of \lemref{propagate:ext} is that the
algorithm has near linear running time for small values of
$U$ and $i$.

Using \lemref{propagate:ext} together with \corref{idiotic}
implies that we can compute a spanning forest for the
``short'' edges of $\Topt(P,L)$ in near linear time.

\begin{lemma}
    Given a set $P$ of $n$ points in the plane, and a set
    $L$ of $n$ lines in the plane. One can compute a
    spanning forest $F$ of $P$, such that the weight of $F$
    is $\leq \eps\Wopt(P,L)/10$. Furthermore, every pair of
    points of $P$ in distance $\Omega( \Wopt(P,L)\eps/(n
    \log^3 n) )$ belong to the same connected components of
    $F$. The running time of this algorithm is $\Ot \pth{ n
    }$.

    \lemlab{start:forest}
\end{lemma}
\begin{proof}
    Using the algorithm of \lemref{rough}, compute in
    $\Ot(n)$ time, a number $M$ such that $\Wopt(P,L) \leq M
    = O(n \alpha(n) \log^2{n} + \Wopt(P,L) \alpha(n) \log
    n)$. In particular, let
    \begin{equation}
        \lshort = \frac{\eps M}{\cAnother n\log^3{n}} \leq
        \frac{\eps}{40n}\Wopt(P,L),
        \eqlab{specify:l}
    \end{equation}
    for $\cAnother$ large enough. On the other hand,
    $\lshort = \Omega( \Wopt(P,L)/ ( U n) )$, where $U =
    O((\log^3{n})/\eps)$.
    
    We now compute a spanning forest for $P$, using
    \lemref{propagate:ext} with $\lshort$ and $U$ as specified and
    $i= 2\rho(l,\eps)$. The running time of this algorithm
    is
    \[
    \Ot \pth{i U n} = \Ot \pth{ \rho(\lshort,\eps) n } = \Ot \pth{
       \frac{\log{n}}{\eps^2} \cdot n } = \Ot\pth{n}.
    \]
    
    Clearly, $F$ has at most $n$ edges, and all the points
    of $P$ in distance $\leq \lshort$ are in the same connected
    component of $F$ by \lemref{good:estimate}.
    
    Furthermore, for any edge $p q$ of $F$, we have that
    with high probability
    $\IDL(p,q) \leq 2(1+\eps)\lshort \leq 4\lshort$ by
    \lemref{good:estimate}. In particular, $\weight(F, L)
    \leq 4n\lshort \leq (\eps/10)\Wopt(P,L)$.
\end{proof}

\lemref{start:forest} implies that we can compute a cheap
spanning forest of $P$ in near linear time that ``captures''
all the light edges of the MST. Next, we can compute the
rest of the edges of the MST using \lemref{propagate:ext}
repeatedly.
\begin{lemma}
    Given a set $P$ of $n$ points in the plane, and a set
    $L$ of $n$ lines in the plane, a parameter $\eps>0$, and
    a spanning forest $F$ of $P$, such that every pair of
    points of $P$ in distance $\leq l$ belong to the same
    connected components of $F$, where $l = \Omega(
    \Wopt(P,L)\eps/(n \log^3 n) )$. Then, one can compute a
    spanning forest $F'$ of $P$ such that all the points of
    $F$ in distance $\leq 2l$ belong to the same connected
    component of $F'$.  The forest $F'$ can be computed in
    $\Ot \pth{ n}$ expected time.
    
    \lemlab{round:forest}
\end{lemma}

\begin{proof}
    We use the same algorithm of \lemref{start:forest}, with
    the modification that when calling to the algorithm of
    \lemref{propagate:ext}, we pass on $F$, such that the
    algorithm ignore generated edges that belong to the same
    connected component of $F$. It is again clear, that only
    edges of length between $l$ and $2 (1+\eps)l$ would be
    added to the spanning forest. The exact details of how
    to specify $U$ and $i$ are similar to
    \lemref{start:forest}, and are omitted.
\end{proof}

Our algorithm for computing the MST works by using
\lemref{start:forest}. This results in a spanning forest
$F_0$ of the points of $P$, and a value $\lshort$ as
specified by \eqref{specify:l}. We now use
\lemref{round:forest} repeatedly $O(\log{n})$ times, in the
$i$-th iteration handling distances between $2^{i-1}\lshort$
to $2 \cdot 2^i \lshort (1+\eps)$, for $i=1, \ldots,
O(\log{n})$), till we handle all distances $\leq n$. Namely,
in the $i$-th iteration, we compute a spanning forest $F_i$
of all points in distance $\leq 2^i\lshort$ from each other
using \lemref{round:forest} using $F_{i-1}$ as our
``starting'' spanning forest.

Clearly, the expected running time of the resulting
algorithm is $\Ot \pth{ n }$.  What is not clear, is that
the resulting MST is indeed an $\eps$-approximate MST.

\begin{lemma}
    With high probability, the tree $T$ computed by the above
    algorithm is an $\eps$-MST of $P$ in $\Arr(L)$.
\end{lemma}

\begin{proof}
    All the edges generated by the algorithm of
    \lemref{start:forest}, in the first stage of the
    algorithm, have total weight $\leq (\eps/20) \Wopt(P,L)$
    with high probability.
  
    Let $\Topt(P,L)$ be the optimal spanning tree. If $T$ is
    not an $\eps$-approximate MST, then $\weight_{\IDL}(T)>
    (1+\eps)\weight_{\IDL}(\Topt)$. In particular, there
    must be an edge of $\Topt$ which its insertion into $T$
    would results in substantially lightly spanning tree.
    Formally, for an edge $e$, let $T(e)$ be the tree
    resulting from $T$ by inserting $e$ into $T$, and
    removing from $T$ the heaviest (according to $\IDL$)
    edge on the new cycle that was created, and let
    $\out(T,e)$ denote this ``ejected'' edge.
    
    Arguing as in the proof of \lemref{approx:mst}, it must
    be that there exists an edge $\phi=p q$ of $\Topt$ such that
    \[
    (1+\eps)\IDL(\phi) < \IDL( \out(T,\phi) ),
    \]
    and $\IDL(\phi) > \Wopt(P,L)/(20n)$.
    
    Let $i$ be the index such that $2^{i-1} \lshort \leq
    \IDL(\phi) \leq 2^i \lshort$. With high probability, we
    know that after the $i$-th iteration $p$ and $q$ are in
    the same connected component of $F_i$. Assume that $p$
    and $q$ were not in the same connected component of
    $F_{i-1}$ (the other case is easier and as such is
    omitted).
    
    Let $T''$ be the spanning forest maintained by the
    algorithm just after $p$ and $q$ were present in the
    same connected component.  With high probability, for
    any edge $e''$ of $T''$, we have $\IDL(e'') \leq
    (1+\eps)\IDL(\phi)$, since the random sample $R_i$ we
    used in the $i$-iteration is $(2^{i-1}
    \lshort,\eps)$-approximation to $\IDL$. 
    
    But then, it is not possible that the algorithm added
    $\out(T,\phi)$ to the spanning tree $T''$, as all the
    edges on the cycle in $T'' \cup \brc{\phi}$ are lighter
    than $(1+\eps)\IDL( \phi)$. A contradiction.
\end{proof}

We summarize our result:
\begin{theorem}
    Given a set $P$ of $n$ points in the plane, $L$ a set of
    $n$ lines, and $\eps > 0$ a parameter. Then one can
    compute a spanning tree $T$ of $P$, in $\Ot \pth{ n }$
    expected time, such that $\weight(T, L) \leq
    (1+\eps)\Wopt(P,L)$. The result is correct with high
    probability.
\end{theorem}

%----------------------------------------------------------------
%----------------------------------------------------------------

\section{Approximation Algorithms for the Intersection
   Metric via Embeddings}

\seclab{embed}

Let $P=\brc{p_1, \ldots, p_n}$ be a given set of $n$ points,
and $L = \brc{l_1, \ldots, l_m}$ be a set of $m$ lines,
where $m= n^{O(1)}$.  As mentioned earlier, the metric
$\IDL$ is computationally cumbersome. One possible way to
overcome this problem, is to embed this metric into a more
convenient metric (while introducing a small distortion
error).

In this section, we show a somewhat weaker result. We show
how to embed the points of $P$ into $O(\log^\EmbedDim
n)$-dimensional space in $\Ot(n+m+n^{2/3}m^{2/3})$ time, so
that a specific distance gap in the crossing metric, is
mapped to a corresponding gap in the target space.

We first observe that the crossing distance between two
points $p$ and $q$, can be computed by interpreting this
distance as a Hamming distance on the hypercube in $m$
dimensions induced by the lines. Namely, each line $l$
contribute a coordinate --- a point gets a '1' in this
coordinate if it is on one side of $l$, and a '0' if it is
on the other side of $l$. Formally, let $l^+$ denote the
open half-plane defined by a line $l$ that contains the
origin, and $l^-$ denote the other open plane.  For a point
$p \in \Re^2$, let $\vL(p) = (b_1, \ldots, b_m)$ be a
$m$-bit vector so that $b_i=1$ {\bf iff} $p \in l_i^+$.  It
is easy to verify that $\IDL(p,q) = d_H(\vL(p), \vL(q))$,
where $d_H$ is the Hamming distance.

\remove{
On this mapped set, we can now deploy several approximation
algorithms for points in high-dimension. However, all those
algorithms first need to read all their input, which
requires $\Omega(nm)$ time. A standard technique to reduce
the dimension of the input (and thus its size), while
preserving distances between points, is to use dimension
reduction techniques \cite{jl-elmih-84,im-anntr-98}.  We
next show how one performs a (somewhat restricted) dimension
reduction in an implicit way, by using the underlining
geometry in $o(m n)$ time.
}

\begin{defn}
    Let $R \subseteq L$, let $f_R:\Re^2 \rightarrow \ZZ$ be
    the mapping that maps a point $p$ in the plane to its
    face ID in the arrangement $\Arr(R)$. Formally, we
    assign for each face in the arrangement $\Arr(R)$ a
    unique integer (say, and integer between $1$ and
    $O(|R|^2)$). The mapping $f_R$ maps a point $p$ in the
    plane to the integer identifying the face that contains
    $p$. (Note, that is does not uniquely define
    $f_R(\cdot)$ as we did not specify how we assign the IDs
    to the faces.)
    
    For a set $\R = (R_1, \ldots, R_\mu)$ of subsets of $L$,
    let $f_\R:\Re^2 \rightarrow \ZZ^\mu$ be the mapping
    $f_\R(p) = ( f_{R_1}(p), f_{R_2}(p), \ldots,
    f_{R_\mu}(p))$. For two points $p,q \in \Re^2$, let
    $d_H(f_\R(p),f_\R(q))$ be the Hamming distance between
    $f(p)$ and $f_\R(q)$. Namely, this is the number of
    coordinates, where the two vectors $f_\R(p)$ and
    $f_\R(q)$ disagree.

    One can view $f_\R$ as an embedding of the crossing
    metric $\IDL$ to the Hamming space $\ZZ^\mu$. 
\end{defn}

\begin{lemma}
    Given a set $P$ of $n$ points in the plane, a set $L$ of
    lines in the plane, a parameter $\eps > 0$ and a
    parameter $r$. One can compute a set $\R$ of $\mu$
    subsets of $L$, such that for the embedding $f_\R:\Re^2
    \rightarrow \ZZ^\mu$, we have that, with high
    probability, for any $p,q \in P$ it holds:
    \begin{itemize}
        \item If $\IDL(p,q) \leq r$, then $d_H(f(p), f(q))
        \leq M$, 
        \item If $\IDL(p,q) \geq (1+\eps)r$ then $d_H(f(p),
        f(q)) \geq (1+\eps)(1-a/\log{n})M$,
    \end{itemize}
    where $M$ and $a$ are appropriate constants and $\mu
    =O(\log^4 n)$.

    \lemlab{good:embed}
\end{lemma}
\remove{
In the following, we restrict ourselves to the case where
only distances in a certain range are approximately
preserved by the embedding.  Namely, for a prescribed
parameters $r > 0$, $\eps > 0$ we describe a mapping
$f(\cdot)$ so that if a pair of points $p,q$ is in distance
$\leq r$, then it is mapped (with high probability) into a
pair $f(p),f(q)$ having distance $\leq M$, and if $p,q \geq
(1+\eps)$, then the pair $f(p),f(q)$ are in distance
$\geq(1+\eps')M$, where $M$ is an appropriate constant, and
$\eps, \eps'$ are of the same up to the factor of
$(1+O(1)/\log n)$.

  In this way approximate nearest neighbor
in the original space with error $(1+\eps)$ is be reduced
the $(1+\eps')$-approximate nearest neighbor in the
resulting Hamming space.  For the purpose of using the
nearest neighbor algorithms of~\cite{im-anntr-98} this
``threshold embedding'' is sufficient,
see~\cite{im-anntr-98} for details.
}

\begin{proof}
    For sake of simplicity of exposition, we assume that $m
    / r \geq \log{n}$, where $m=|L|$.  If this is not
    correct, we can add ``fictitious'' lines to $L$ that have
    all the points of $P$ on one side of them. If we pick
    such a line to a set of $\R$, we can ignore it when we
    compute the face IDs. 
    
    For a parameter $\alpha$ to be specified shortly, let
    $k= \alpha m/r$, $R$ be a sample of $k$ lines out of $L$
    (performed with replacement), and let $p,q$ be two
    points of $P$.  Let $\rho = \IDL(p,q) /n$.  The
    probability that $p,q$ will be in two different faces of
    $\Arr(R)$ is
    \[
    U(\rho) = 1 - (1-\rho)^k,
    \]
    as this is the probability that not all the lines will
    miss the segment connecting $p$ and $q$.
    
    Our target is to approximate the value of $U(\rho)$ so
    we could decide whether $p,q$ are close or far. Indeed,
    if $U(\rho) \geq U( (1+\eps)r/m )$ then $\IDL(p,q) \geq
    (1+\eps)r$, and if $U(\rho) \leq U( r/m )$ then
    $\IDL(p,q) \leq r$.
    
    To do so, we generate a set of subsets $\R = (R_1,
    \ldots, R_{\mu})$, by random sampling as described
    above, where $\mu$ would be specified shortly.  Now we
    consider the quality of the distance approximation
    provided by the embedding\footnote{A similar analysis
       (in the context of Hamming spaces) appeared already
       in~\cite{i-drtpp-00}; in our case, however, we have
       to put more care into the analysis, since we want
       $\eps$ and $\eps'$ to be very close.}.  Let
    $X(p,q)$ denote the random variable which is the number
    of arrangements of $\Arr(R_1), \ldots, \Arr(R_\mu)$ that
    have $p,q$ in different faces.  Note, that $X(p,q)$ is
    equal to the Hamming distance between $f_\R(p)$ and $f_\R(q)$,
    and it thus the distance between the images of $p$ and
    $q$ in the new space.  Clearly, as $\mu$ tends to
    infinity, $X(p,q)/\mu$ tends to $U(\rho)$. Using
    Chernoff inequality, we can quantify the quality of
    approximation provided by $\mu$.  Specifically, let
    $\plow=U(r/m)$ and $\phigh=U((1+\eps)r/m)$; in the
    following we will make sure that $\phigh<1/2$.  Then,
    from the Chernoff bound~\cite{mr-ra-95,mps-lpvaa-98} it
    follows that for any $\alpha>0$ if $\mu=C \frac{\log
       n}{\plow \alpha^2}$ for some constant $C$, then with
    high probability:
    \begin{itemize}
        \item if $\IDL(p,q) \le r$ then $X(p,q)/\mu \le
        \plow(1+\alpha)$
    
        \item if $\IDL(p,q) \ge r(1+\eps)$ then $X(p,q)/\mu
        \ge \phigh(1-\alpha)$
    \end{itemize}    
    Therefore, the mapping $f_\R$ converts the distance gap
    $r:(1+\eps)r$ into the gap $\plow(1+\alpha)\mu :
    \phigh(1-\alpha)\mu$.  We next fine tune $k$ (the size
    of each sample) so that the resulting gap will be as
    large as possible. (Intuitively, the larger the target
    gap is, the easier it is to detect it in later stages.)
    Therefore, in the following we focus on finding $k$ such
    that the ratio
    \[
    \Delta = \frac{\phigh(1-\alpha)\mu}{\plow(1+\alpha)\mu}
    \]
    is as large as possible.  To this end, we observe that
    \begin{eqnarray*}
        \plow &=& U\pth{\frac{r}{m}}=1-\pth{1-\frac{r}{m}}^k
        \leq  1 - e^{-r k/m}\pth{ 1- \frac{(r k/m)^2}{k}} 
        =  1 - e^{-\alpha}\pth{ 1- \frac{\alpha^2}{k}} \\
        &\leq&
        1 - e^{-\alpha}\pth{ 1 - \alpha^2}
        \leq \alpha^2 + (1 - e^{-\alpha}) \pth{ 1 -
        \alpha^2}
        \leq \alpha^2 + \alpha \pth{ 1 -
           \alpha^2}
        \leq \alpha(1+\alpha)
    \end{eqnarray*}
    since $\displaystyle \pth{ 1 - \frac{t}{n}}^{n} \geq
    e^{-t} \pth{ 1 - \frac{t^2}{n}}$~\cite{mr-ra-95},
    $k=\frac{\alpha m}{r}$, and $x \geq 1 -e^{-x}$.
    Furthermore,
    \begin{eqnarray*}
        \phigh  &=&  U((1+\eps)r/m) =1-(1-(1+\eps)r/m)^k
        \geq 1-e^{-(1+\eps)r k/m}
        = 1-e^{-(1+\eps)\alpha}\\
        & \geq & (1+\eps)\alpha - ((1+\eps)\alpha)^2 
         \geq  (1+\eps)\alpha(1 - (1+\eps)\alpha)
    \end{eqnarray*} 
    since $(1-t/n)^{n} \leq e^{-t}$~\cite{mr-ra-95} and 
    $1-e^{-x} \ge x-x^2/2 \geq x - x^2$.
    
    Therefore
    \begin{eqnarray*}
        \frac{\phigh}{\plow} &\geq&
        \frac{(1+\eps)\alpha(1 -
           (1+\eps)\alpha)}{\alpha(1+\alpha)}
        \geq (1+\eps)(1 - (1+\eps)\alpha)(1-\alpha)
        \geq (1+\eps)(1 - (2+\eps)\alpha),        
    \end{eqnarray*}
    since $1/(1+x) \geq (1-x)$.  Thus, if we set $\alpha$ to
    be $1/\log n$, then the distance gap becomes (at least)
    \[
    \Delta = \frac{\phigh(1-\alpha)\mu}{\plow(1+\alpha)\mu} \geq 
    (1+\eps)(1-(2+\eps)\alpha)(1-\alpha)^2 \geq
    (1+\eps)\pth{1 - \frac{a}{\log{n}}},
    \]
    where $a$ is an appropriate constant.  Also, note that
    the resulting value of $\plow$ is
    \[
    \plow =
    1-(1-r/m)^k  
    \geq 1 - e^{-p_0k} = 1 -e^{-\alpha} \geq \alpha - \alpha^2/2 =
    \Omega(1/\log n)
    \]
    and $\mu=(C\log{n})/(z\alpha^2) = C\log^2{n}/\alpha^2 =
    O(\log^4 n)$.  Finally, since $m/r \geq \log{n}$, we
    have that $k= \alpha(m/r) = (1/\log{n}) (m/r) \geq 1$
    (i.e., the sample size $k$ is at least $1$).
\end{proof}    

\begin{lemma}
    Given a set $P$ of $n$ points, and a set $L$ of $m$
    lines, one can compute the function $f_\R(\cdot)$, of
    \lemref{good:embed}, for all the points of $P$ in
    $\Ot( (m^{2/3}n^{2/3} + m + n))$ expected time.
\end{lemma}

\begin{proof}
    We have to compute for each point of $P$ the face that
    contains it in each of the arrangements $\Arr(R_1),
    \ldots, \Arr(R_\mu)$, where $\mu = O( \log^4 n )$. Or
    alternatively, compute all the faces of $\Arr(R_1),
    \ldots, \Arr(R_\mu)$ that contains points of $P$. For a
    single arrangement $A_i$ this can be done in \linebreak
    $O(m^{2/3}n^{2/3} \log^{2/3}(m/\sqrt{n}) + (m +
    n)\log{m})$ expected time \cite{ams-cmfal-98}. Since
    there are $\mu$ coordinates (i.e., arrangements), the
    result follows.
\end{proof}

Thus, we showed how to embed $\IDL$ into $\mu$-dimensional
Hamming space $\Sigma^{\mu}$ in $\Ot(n+m+n^{2/3}m^{2/3})$
time, mapping a $(1+\eps)$ gap between close and far points
into a gap of size $(1+\eps)(1-O(1)/\log{n})$, where $\mu =
O(\log^4 n)$ and $\Sigma \subseteq \ZZ$ is the set of face
labels we use (i.e., $|\Sigma| = O(m^2)$.  By using standard
embedding techniques (e.g.  see~\cite{kor-esann-00}) we can
embed the Hamming space $\Sigma^{\mu}$ into $\{0,1\}^D$ with
$D=O(\mu \log |\Sigma| \log^2 n) = O(\log^{6}{n} \log{m})$,
preserving the gap up to another factor $(1-O(1)/\log n)$.
This gives an embedding of $\IDL$ into $D=O(\mu \log m
\log^2 n)$-dimensional binary Hamming cube, with error
$(1-O(1)/\log n)$.  Thus it is sufficient for us to maintain
$c$-nearest neighbor in $\{0,1\}^D$ where $c=(1+\eps)
(1-O(1)/\log n)$, which takes
$\Ot(n^{1/c})=\Ot(n^{1/(1+\eps/2)})$ time per operation
\cite{im-anntr-98}.

We conclude:
\begin{theorem}
    By performing a $\Ot(n+m+n^{2/3}m^{2/3})$-time
    preprocessing, one can reduce the problem of maintaining
    dynamic $(1+\eps)$-approximate nearest neighbor for any
    $n$-point crossing metric over $m$ lines, to the problem
    of maintaining dynamic $(1+\eps)(1-O(1)/\log
    n)$-approximate nearest neighbor in Hamming space with
    $O(\log^\EmbedDim n)$ dimensions (assuming $m=
    n^{O(1)}$).  The latter can be solved in
    $\Ot(n^{1/(1+\eps/2)})$ time per operation.
\end{theorem} 

\remove{
   \subsection{Embedding of the Crossing Metric over $\Re^d$}
   
   In this Section, we extend the methods from the previous
   section to the crossing metric defined by
   $d-1$-dimensional hyperplanes in $\Re^d$, for any fixed
   $d \ge 2$.  To this end, it is sufficient to design an
   efficient procedure, which given a set of $n$ points
   $p_1, \ldots, p_n$ and $m$ hyperplanes $H_1, \ldots,
   H_m$, assigns a symbol $a_i \in \Sigma$ to each $p_i$ in
   such a way that $a_i \neq a_j$ iff there exists $H_k$
   which separates $p_i$ from $p_j$.  Unfortunately, the
   idea from the previous section does not give subquadratic
   time algorithm for $d>2$, since even in $d=3$ the
   complexity of $n$ arrangement cells from an arrangement
   formed by $n$ planes could be $\Omega(n^2)$.
   Fortunately, for our purpose, we do not need to compute
   the actual cells containing $p_i$s; rather, it is just
   sufficient to find {\em labels} of those cells.

   The algorithm for finding the labels is based on {\em
      partition trees} by \matousek{}~\cite{m-ept-92}, which
   are defined as follows.  }

\subsection{Embedding of the Crossing Metric over $\Re^d$}

In this Section, we extend the methods from the previous
section to the crossing metric defined by
$(d-1)$-dimensional hyperplanes in $\Re^d$, for any fixed $d
\ge 2$.  To this end, it is sufficient to design an
efficient procedure, which given a set of $n$ points $P=p_1,
\ldots, p_n$ and a set of $m$ hyperplanes $\HX = \brc{H_1,
   \ldots, H_m}$, assigns a symbol $\sigma_i \in \Sigma
\subset \ZZ$ to each $p_i$ in such a way that $\sigma_i \neq
\sigma_j$ iff there exists $H_k$ which separates $p_i$ from
$p_j$.  Unfortunately, the idea from the previous section
does not give subquadratic time algorithm for $d>2$, since
even in $d=3$ the complexity of $n$ cells in an arrangement
formed by $n$ planes could be $\Omega(n^2)$.  Fortunately,
for our purpose, we do not need to compute the actual cells
containing $p_i$s.  Rather, it is just sufficient to find
the {\em labels} for those cells, or more specifically, a
function $h: P \to \Sigma$ such that $h(p)=h(q)$ iff $p$ and
$q$ belong to the same arrangement cell.

Abusing notations, we denote by $H_k(p)$ the function
returning $1$ if $p$ lies on one side of $H_k$ and zero
otherwise. We use the following hashing function
\[
h(x)= \pth{\sum_i a_i H_i(x)},
\]
where $a_1 \ldots a_m$ are independent and identically
distributed random variables with uniform distribution over
$\brc{0, \ldots ,n^c}$, where $c$ is a constant to be
specified shortly.  Note, that if $p,q \in \Re^d$ lie in two
different full-dimensional faces of $\Arr(\HX)$, then, as
noted above, there must be a hyperplane $H_k \in \HX$, so
that $H_k(p) \neq H_k(q)$, and say that $H_k(p) = 1$. That
is, $h(p) = h'(p) + a_k$ and $h(q) = h'(q)$, where $h'(x) =
\sum_{i\neq k} a_i H_i(x)$. Since the $a_i$ were picked
independently, it follows that $h(p)=h(q)$ only if $h'(p) -
h'(q) = a_k$. But the probability of that to happen is
$1/n^c$. We conclude, that the probability of two points
belonging to two different faces to be mapped to the same
value by $h(\cdot)$ is $1/n^c$. Thus, since we have $O(n^2)$
pairs of points to consider in our algorithm, it follows
that the probability of the hashing to fail is $n^{2-c}$
which can be made to be arbitrarily small by picking $c$ to
be large enough.

Namely, we associate a weight $a_i$ with each half-space
induced by a hyperplane $H_i$. For each point $p_j$, we
compute the total weight of all the half-spaces that contain
it, and all the points having the same total weight are
associated with the same label. Computing the weight of a
point $p_j$ falls into the class of problems known as
intersection-searching \cite{a-rs-97}. In particular, one
can construct a data-structure in $O(m^{1+\delta})$ time, so
that one can answer intersection-searching queries in $O(
(n/m^{1/d}) \log^{d+1} n )$ time, where $\delta >0$ is
arbitrarily small constant. As the algorithm needs to perform
a linear number of such queries, we set $m= n^{2d/(d+1)}$.
Thus, the algorithm computes the required labels in
$O(n^{2d/(d+1) + \delta})$ time.
We conclude:
\begin{theorem}
    By performing a $O(n^{2d/(d+1)+\delta})$-time
    preprocessing, where $\delta >0$ is arbitrary constant,
    one can reduce the problem of maintaining dynamic
    $(1+\eps)$-approximate nearest neighbor for any
    $n$-point crossing metric over $n$ hyperplanes in
    $\Re^d$, to the problem of maintaining dynamic
    $(1+\eps)(1-O(1)/\log n)$-approximate nearest neighbor
    in Hamming space with $O(\log^\EmbedDim n)$ dimensions.

    \theolab{reduction}
\end{theorem}

\begin{remark}
    Note, that the constants in the bounds of Theorem
    \ref{theo:reduction} depend exponentially (or worse) on
    the dimension $d$.
\end{remark}

\begin{remark}
    As indicated in the introduction, having such a
    embedding, enable one to use a large collection of
    subquadratic approximation algorithms for the
    intersection metric, including dynamic amortized
    $\Ot(n^{4/3} + n^{1+1/c})$-time (for $d=2$)
    $c$-approximation algorithms for bichromatic closest
    pair~\cite{e-demst-95} and $\Ot(n^{4/3} +
    n^{1+1/c})$-time algorithms for: $c$-approximate
    diameter and discrete minimum enclosing ball
    \cite{giv-rahdp-01}, $O(c)$-approximate facility
    location and bottleneck matching~\cite{giv-rahdp-01}.
    Similar (i.e., subquadratic time) results hold for any
    $d>2$.
\end{remark}

\subsection{Computing an MST Using the Embedding}

We next describe how to use the embedding described in the
previous two sections, for getting an
$(1+\eps)$-approximation algorithm for the MST under
crossing metric. Note that everything described in this
section is well known \cite{im-anntr-98}, and we provide it
only for the sake of completeness. Also, the resulting
algorithm is slower in the planar case than the algorithm of
Section \ref{sec:speedup}.

Computing the minimum spanning tree under the intersection
metric, using the Kruskal's algorithm, boils down to
maintaining the bichromatic nearest-neighbor pair (under
the intersection metric) between two sets $P_1, P_2
\subseteq P$, under insertions and deletions. A consequence
of Eppstein result \cite{e-demst-95} is the following:

\begin{theorem}[\cite{e-demst-95}]
    Given a dynamic data-structure for nearest-neighbor
    queries, where each insertion / deletion / query operation
    takes $T(n)$ time, then one can compute the MST in
    $O(n T(n)\log^2 n)$ time.
\end{theorem} 

It is easy to verify that if we get a
$(1+\eps)$-approximation to the MST if we use an
$(1+\eps)$-approximate dynamic nearest-neighbor
data-structure (Eppstein, personal communication, 1999).

Namely, we need a data-structure that support dynamic
approximation nearest-neighbor queries.  After applying the
embedding described above, we use the $\eps'$-PLEB
data-structure of \cite{im-anntr-98} to maintain a
$(1+\eps')$-approximate nearest neighbor in the embedded
space. Specifically, we construct an $\eps$-PLEB in the
embedded points. In this way, we obtain an $\eps$-PLEB for
our original points (i.e., we embedded a gap to a gap, so
that a close point in the embedded space, corresponds to a
close point in the crossing metric) data-structure that for
a query $p$ return us a point of $q \in P$ so that
$\IDL(p,q) \leq (1+\eps)r$, if there exits a point $q^* \in
P$ so that $\IDL(p,q^*) \leq r$.

Thus, by constructing $\log_{1+\eps}n$ such data-structures,
we can use binary search on those data-structures to find
and $(1+\eps)$-approximate nearest neighbor to a query
point.  Namely, this data-structure can be used to answer
approximate nearest neighbor queries for the intersection
metric.  For the whole scheme to work, we need those
data-structures to be dynamic; i.e., support insertions and
deletions of points.  Fortunately, the only part of the
algorithm that needs to be dynamic is the second stage that
uses the data-structure of \cite{im-anntr-98} which is
dynamic.
 
We conclude:
\begin{theorem}
    Given a set $P$ of $n$ points in the plane, and a set
    $L$ of $n$ lines, one can compute in $\Ot \pth{ n^{4/3}
       + n^{1+ 1/(1+\eps)} }$ time, a spanning tree of $P$
    of weight $\leq (1+\eps)\Wopt(P,L)$. The result returned
    by the algorithm is correct with high probability.  For
    $d>2$ dimensions, such an MST can be approximated in
    $\Ot \pth{ n^{2d/(d+1) + \delta} + n^{1+ 1/(1+\eps)} }$
    time, where $\delta>0$ is an arbitrary constant.
\end{theorem}

%----------------------------------------------------------------
%----------------------------------------------------------------
%----------------------------------------------------------------
%----------------------------------------------------------------
\section{Conclusions}
\seclab{conc}

We presented the first $(1+\eps)$-algorithm for
approximating the minimum spanning tree under the crossing
metric in the plane.  We also presented a subquadratic time
approximation algorithms for a variety of other problems,
obtained by embedding the crossing metric into higher
dimensional space.  The techniques used in our paper seems
to be new to low-dimension computational geometry, and we
believe that they might be useful for other problems in
computational geometry.

There are several interesting open problems for further
research:
\begin{itemize}
    \item Can the result be extended to other cases:
    segments or arcs instead of lines? 

    \item Can a similar approximation algorithm be found
    for the case of minimum weight triangulation under the
    crossing metric? 
\end{itemize}

\subsection*{Acknowledgments}
 
The authors wish to thank Pankaj Agarwal, Boris Aronov and
Micha Sharir for helpful discussions concerning the problems
studied in this paper and related problems.

%-------------------------------------------------------------------------
% Bibliography 
%-------------------------------------------------------------------------
\bibliographystyle{salpha} 
\bibliography{shortcuts,geometry}

%-------------------------------------------------------

\appendix
\section{A Rough Approximation to the Weight of the
   MST in Near Linear Time}
\seclab{fast:approx}

In this appendix, we show how to approximate the weight of
the minimum spanning tree up to roughly a factor of
$O(\alpha(n)\log{n})$ if its weight is at least linear.  In
Section \ref{sec:speedup}, we presented a near linear time
algorithm for $(1+\eps)$-approximation for the minimum
spanning tree, that relies on this approximation algorithm.

Underlining the approximation algorithm, is the observation
that an MST for a random sample of the lines of $L$ provides
a rough approximation to the weight of the MST of $L$.
If the weight of the MST of the sample is near linear,
we can approximate it up to a $O(\alpha(n)\log{n})$, using
the following algorithm.

\begin{lemma}
    Given a set $R$ of $r$ lines, $P$ a set of $n$ points,
    and $W$ a prescribed parameter, one can decide whether
    $\Wopt(P,R)$ is large; namely, $\Wopt(P,R) = \Omega( (r
    + n + W) \alpha(n)\log{n} )$.  The algorithm takes $O(
    (r + n + W) \alpha(n)\log^2{n} )$ expected time.
    Furthermore, if $\Wopt(P,R) \leq W$, the algorithm will
    report that its weight is large with probability at most
    $n^{-c}$, where $c$ is an appropriate constant.
    
    \lemlab{brute:estimate}
\end{lemma}

\begin{proof}
    Use the algorithm of Theorem~\ref{theo:hs} and execute
    it $O(\log{n})$ times on $P$ and $R$. If the running
    time of the $i$-th execution of the algorithm exceeds
    $\Omega( (r + n + W)\alpha(n) \log{n})$ abort it, and
    move on to the next execution. If $\Wopt(P,R) \leq W$,
    then the algorithm of \cite{hs-oplpa-01-dcg} provides a
    spanning tree of expected weight $O( (r+n+ W)\alpha(n)
    \log{n})$ with the same bound on the expected running
    time. Thus, if in $O(\log{n})$ executions the algorithm
    returns always that $\Wopt$ is large, we can conclude
    that with probability $\geq 1 - n^{-c}$ the weight of
    $\Wopt(P,R)$ is not $\leq W$.
\end{proof}

\lemref{brute:estimate} shows that we can
approximate the weight of the MST in near linear time if its
weight is near linear. However, if it is heavier, we will
use random sampling to keep the running time under control.

Let $R \subseteq L$ be a random sample of lines out of $L$,
where each line is picked independently with probability
$r/n$.  Clearly, the probability of an intersection point
$u$ (between a connected set $\gamma$ and a line of $L$), to
be present in $\Arr(R)$ is $r/n$ (this is the probability
that the line of $L$ passing through $u$ will be chosen to
be in the random sample).
 
\begin{defn}
    For a curve $\gamma$, and a set of lines $L$, let
    $\weight(\gamma,L)$ denote the {\em weight} of $\gamma$
    in the arrangement $\Arr(L)$. This is the number of
    intersections of $\gamma$ with the lines of $L$.
\end{defn}

\begin{lemma}
    Let $R$ be a sample of lines of $L$ (chosen as described
    above), then with high probability:
    \[
    \Wopt(P,L) \leq \frac{n}{r} \pth{ c_0 n \log{n} + 2
       \Wopt(P,R)},
    \]
    and with probability $\geq 0.9$ we have $\frac{n}{r}
    \cdot \frac{\Wopt(P,R)}{10} \leq \Wopt(P,L)$, where $c_0$ is an
    appropriately large constant.
    \lemlab{wopt:sample}
\end{lemma}

\begin{proof}
    Let $\Topt^L = \Topt(P,L)$, and let $W_R =
    \weight(\Topt^L, R)$ be the weight of $\Topt^L$ under
    the crossing metric of $R$. Clearly, $E[ W_R] =
    \Wopt(P,L)\frac{r}{n}$. Thus, we know that with
    probability $\geq 0.9$ we have $W_R \leq 10 \Wopt(P,L)
    \frac{r}{n}$ (by Markov inequality), and with
    probability $\geq 0.9$, we have that $\displaystyle
    \Wopt^R = \Wopt(P,R) \leq W_R \leq 10 \Wopt(P, L)
    \frac{r}{n}$.

    Let $p,q \in P$ be two points, and let $X_{p q}$ be
    the distance between $p,q$ in the arrangement $\Arr(R)$.
    If the distance between $p,q$ is large, that is $U =
    \IDL(p,q) \geq c_0 (n/r) \log{n}$ (where $c_0$ is a
    large enough constant), then one can show using Chernoff
    inequality, that with high probability, we have:
    \[
    \frac{U}{2} \leq X_{p q} \frac{n}{r} \leq 2 U.    
    \]
    
    On the other hand, by the above argument, each edge $e
    =p q$ of $\Topt^R = \Topt(P,R)$ either intersects at most
    $c_0 (n/r)\log{n}$ lines of $L$, or alternatively, the
    number of lines of $L$ intersected by $e$ is smaller
    than $2(n/r)X_e$, where $X_e$ is the number of lines of
    $R$ that $e$ intersects.  Thus, with high probability,
    we have
    \begin{eqnarray*}
        \Wopt(P,L) &\leq & \weight( \Topt^R, L) = \sum_{e =
           p q \in \Topt^{R}} \IDL(p,q)
        \leq \sum_{e \in
           \Topt^{R}} \pth{ c_0 \frac{n}{r}\log{n} +
           2X_e\frac{n}{r}}\\
        &=& c_0 \frac{n^2 \log{n}}{r}
        + \Wopt(P,R) \frac{2n}{r}.
    \end{eqnarray*}
\end{proof}

\begin{remark}
    We can make both probabilities in \lemref{wopt:sample}
    large by repeating the experiment $O(\log{n})$ times,
    and picking the smallest $W(P,R)$ computed. With high
    probability, we have
    \[
    \frac{n}{r} \cdot \frac{\Wopt(P,R)}{10}
    \leq \Wopt(P,L) \leq
    \frac{n}{r} \pth{ c_0 n \log{n} + 2 \Wopt(P,R)}.
    \]
    In particular, if $\Wopt(P,R) > c_0 n \log{n}$, we get
    that $3\Wopt(P,R)\frac{n}{r}$ is a constant factor
    approximation to $\Wopt(P,L)$.
\end{remark}

\begin{lemma}
    Let $r$ be a prescribed parameter, and $\Wopt =
    \Wopt(P,L)$.  Then, an algorithm can decide whether
    \begin{itemize}
        \item $\Wopt$ is small -  namely $\Wopt \leq
        \frac{10c_0 n^2\log{n}}{r} $.

        \item $\Wopt$ is large - $\Wopt = \Omega( 
        \frac{n^2}{r} \alpha(n)\log^2{n})$.
        
        \item $\Wopt$ is in between. Any of the two above
        answers are valid.
    \end{itemize}
    The algorithm takes $O( n \alpha(n)\log^4{n})$ time, and
    returns a correct result with high probability.
    
    \lemlab{fine:estimate}
\end{lemma}

\begin{proof}
    We pick $m = O(\log{n})$ samples $R_1, \ldots, R_m$ by
    picking each line with probability $r/n$ into the
    sample.  For each sample, we check whether $\Wopt(P,R_i)
    \leq 10 c_0 n\log{n}$, using the algorithm of 
    \lemref{brute:estimate}. This will require
    $O(n\alpha(n)\log^{3}(n))$ time for each sample, and
    $O(n\alpha(n)\log^{4}(n))$ overall.
    
    If the algorithm of \lemref{brute:estimate}
    returned {\em not large} for any sample $R$, we know
    that $\Wopt(P,R) = O(n\alpha(n)\log^2{n})$.  And by
    \lemref{wopt:sample}, we know that $\Wopt(P,L)
    = O\pth{ \frac{n^2 \alpha(n)\log^{2}n }{r} }$ with high
    probability.  \remove{On the other hand, if all the
       spanning trees are ``long'' for all the samples, we
       know that $\Wopt(P,L) \geq \frac{n}{10r}\cdot 10 c_0
       n\log{n} = \frac{c_0n^2 \log{n}{r}$ with high
          probability by \lemref{wopt:sample}.  }}
\end{proof}
 
Now, we can perform a binary search to approximate the
weight of $\Wopt(P,L)$.

\begin{lemma}
    One can compute in $O(n\alpha(n)\log^{5}{n})$ time a
    value $M$, so that
    \[
    \Wopt(P,L) \leq M = O(n \alpha(n) \log^2{n} + \Wopt(P,L)
    \alpha(n) \log n).
    \]

    \lemlab{rough}
\end{lemma}

\begin{proof}
    Use \lemref{fine:estimate}, set $r_0 = n$.  In
    the $i$-th iteration check whether $\Wopt = \Omega\pth{
       \frac{n^2}{r_i} \alpha(n)\log^2{n}}$, by using the
    algorithm of \lemref{fine:estimate}. If it is,
    we set $r_{i+1} = r_i/2$, and repeat the process.  We
    stop as soon as this check fails.  Then, we know that
    with high probability
    \[   
    \frac{10c_0 n^2\log{n}}{r_{i-1}} 
    \leq 
    \Wopt(P, L) = O\pth{ \frac{n^2 \alpha(n)\log^{2} n
    }{r_i} } = M,
    \]
    implying that $M$ is the required approximation.
\end{proof}

\begin{remark}
    Note, that if algorithm of \lemref{rough} stops after
    the first iteration, then $\Wopt = O(n
    \alpha(n)\log^2{n})$. In such a case the approximation
    we get is much worse then logarithmic.  However, this is
    to some extent the easiest case: Without any sampling we
    get a spanning tree of near linear (or sub linear)
    weight.
\end{remark}

\end{document}